# Data augmentation versus noise compensation for x-vector speaker recognition systems in noisy environments

Mohammad MohammadAmini, Driss Matrouf
*Avignon University*
*LIA (Laboratoire Informatique d'Avignon)*
{mohammad.mohammadamini , driss.matrouf}@univ-avignon.fr

*Abstract*—The explosion of available speech data and new speaker modeling methods based on deep neural networks (DNN) have given the ability to develop more robust speaker recognition systems. Among DNN speaker modelling techniques, x-vector system has shown a degree of robustness in noisy environments. Previous studies suggest that by increasing the number of speakers in the training data and using data augmentation more robust speaker recognition systems are achievable in noisy environments. In this work, we want to know if explicit noise compensation techniques continue to be effective despite the general noise robustness of these systems. For this study, we will use two different x-vector networks: the first one is trained on Voxceleb1 (Protocol1), and the second one is trained on Voxceleb1+Voxveleb2 (Protocol2). We propose to add a denoising x-vector subsystem before scoring. Experimental results show that, the x-vector system used in Protocol2 is more robust than the other one used Protocol1. Despite this observation we will show that explicit noise compensation gives almost the same EER relative gain in both protocols. For example, in the Protocol2 we have 21% to 66% improvement of EER with denoising techniques.

## I. Introduction

In noisy environments the performance of speaker recognition systems dramatically drops. The state-of-the-art DNN based approaches for speaker modeling have made speaker recognition systems more robust. In noisy environments with different unseen noises, the performance of these approaches is more robust than their previous statistical generation (i-vector), but they still have poor performance compared to noise free situations. Among DNN speaker recognition systems, x-vectors are the most promising and successful approach.

A number of studies [1, 2] emphasized on the importance of increasing the number of speakers and data augmentation in training x-vector network to make the system more robust in noisy environments. In this research, besides the aforementioned solutions to extract more robust x-vectors, we propose to add a denoising subsystem before scoring the x-vectors. The architecture of the proposed system is presented in Fig 1.

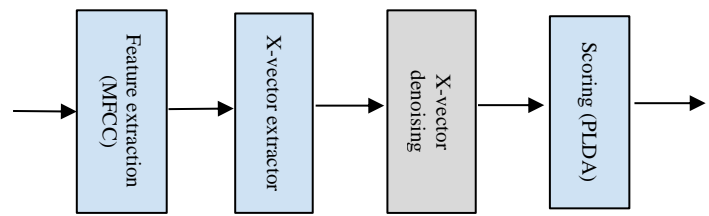

Fig 1. Adding a denoising subsystem to standard x-vector system

Denoising techniques can be used in signal level, feature level, and speaker modeling level. The recent speaker modeling advances (i-vectors or x-vectors) make speaker modeling level suitable for noise compensation. In fact, the data in these spaces are almost Gaussian distributed. Previous studies show the effectiveness of denoising techniques in the i-vector domain [3,4,5,6,7].

Earlier studies in x-vector space mostly focused on x-vector network extractor or data augmentation to improve the performance of speaker recognition system. In [2], it is shown that increasing the number of speakers and using data augmentation makes the x-vector system more robust on noisy and far-field test data. In spite of this fact, the EER obtained with these systems in noisy environments is still much higher than EER obtained in clean conditions. Cycle GANs are another method that are used to transform reverberated log Mel-filter bank features to their clean pairs before training the x-vector network[8]. In [9] it has been shown that speaker embeddings are more robust in far-field test data. But like the additive noise and reverberation there is degradation in comparison to the baseline system. Correlation alignment (CA) algorithm is used for domain adaptation in x-vector space. The CA algorithm tries to minimize the distance between the covariance of the out-of-domain and in-domain x-vectors [10].

There are several attempts that focus on improving the x-vector network. The original x-vector extractor classifies the speakers in the output layer [11]. In [12] a modified speaker embedding network was proposed that classifies the speakers and conditions jointly. The conditions can be continuous (SNR) or discrete (type of noise). This strategy makes the system more robust in different environments. Using non-linear activation function, feature normalization, using CNNs instead of TDNN are among other strategies proposed to improve the performance of x-vector system [13]. In [14] a Gaussian constrained training approach is proposed that impose on x-vectors to have a Gaussian distribution. Using gated convolutional layers instead of time delay layers and gated pooling layer has improved the performance of x-vector system [15]. In [16], a hybrid LSTM and CNN network used for frame level layers and by using multi-level pooling strategy and applying a regularization scheme on embedding layer, the performance of x-vector baseline system was improved. Lie et al, [17] used different variants of large margin SoftMax loss function in x-vector network system.

The above-mentioned modifications of x-vector network don't focus on noise compensation and they try to improve the performance of x-vector system for both noisy and clean environments generally. In [24] we used the statistical i-MAP method to denoise x-vectors. Also, two combinations of i-MAP and denoising autoencoders introduced to deal with additive noise in x-vector space. In this paper we continue our exploration by adding a denoising subsystem to standard x-vector system. This modification is illustrated in Fig.1 . We show that while data argumentation and increasing the number of speakers makes the x-vector system more robust, we can go further and achieve better results by noise compensation techniques. We show that even with large augmented data and a great number of speakers, the noise compensation techniques are effective. To do that, we train two x-vector systems. In the first one, the x-vector network is trained with Voxceleb1 and in the second one the network is trained with a combination of Voxceleb1 and Voxceleb2. In both cases, the train data is augmented with all branches of Musan corpus. We show that in both protocols the relative gain of EER after denoising x-vectors is significant. Hence, denoising techniques even with the availability of huge data is a good solution to increase the robustness of speaker recognition systems.

We use two different denoising techniques. Firstly, we try to find the best denoising autoencoder for x-vector space. Then we propose a stacked denoising autoencoder that deeper autoencoders accepts two set of inputs are used in the input layer. The first one comes from the output of the previous denoising autoencoder and the second one is the difference between noisy x-vectors and the output of the previous denoising autoencoder. Our proposed denoising techniques try to do noise compensation generally, and do not consider a specific kind of noise.

In the following, we introduce the denoising techniques on section II. In section III, the details of experiments setup and protocols are described. Section IV presents the result of the baseline system and denoising techniques.

## II. Denoising techniques

In this section we describe the methods we used to denoise x-vectors. Firstly, we describe the architecture of denoising autoencoder. Then we propose a novel denoising autoencoder named deep stacked denoising autoencoder.

### A. Denoising autoencoder

Denoising autoencoders are among the commonly used noise compensation techniques. The denoising autoencoder is a specific kind of autoencoder that takes the noisy x-vector in the input and creates the clean version at the output. Denoising autoencoder tries to minimize $L(x, f(y))$, in which L is the loss function, $x$ is the clean x-vector, $y$ is the corrupted x-vector and $f(y)$ is the denoised x-vector [18]. Finding a good architecture and its parameters depends on the application and the type of data. In our experiments, we achieved our better results with dense tanh layers in which the number of hidden neurons is greater than the size of x-vector dimensions. The details of parameters and hyperparameters are described in section IV.

### B. Deep stacked denoising autoencoder

In this subsection, we introduce a novel denoising autoencoder named deep stacked denoising autoencoder. In this architecture we have several DAE blocks. The noisy x-vectors fed to the first DAE. The next DAE block receives $X_i$ (the output of the previous block) concatenated with $Z_i = Y - X_{i-1}$ (the difference between noisy x-vectors and the output of the previous block). The stacked DAEs are trained jointly with the stochastic gradient descent optimization algorithm. The architecture of deep stacked DAE is presented in Fig 2. As we can see in the next section, in all cases the deep stacked architecture outperforms denoising autoencoder.

The idea behind this architecture comes from conventional autoencoders that try to copy their input at the output layer. Autoencoders usually are used to create embeddings or they are used for dimensionality reduction. We assume that feeding the difference between noisy x-vectors and the output of the previous denoising autoencoder can find an estimation of the noise information. We did an experiment with autoencoders with Y (noisy x-vector) and Y-X (the exact information of the noise in the x-vectors) in the input to recreate the X in the output layer. We observed that recreated x-vectors are very close to clean x-vectors. So, we infer that feeding information about the noise is helpful in finding more exact denoised x-vectors.

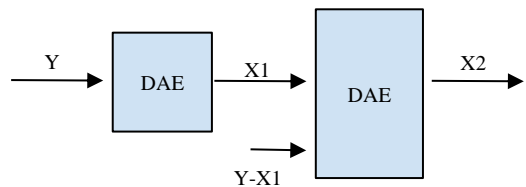

Fig 2. Deep stacked denoising autoencoder

## III. Experiments setup

### A. Corpus

In this subsection the datasets are described briefly:

**Voxceleb:** We used both Voxceleb1 and Voxceleb2. Voxceleb1 includes 100,000 speech files from 1,251 speakers. Voceleb2 has more than 1 million files from 6,000 speakers. The speakers are from different ethnicities with different ages, professions and accents. The utterances have different types of noises including background chatter, laughter, overlapping speech and room acoustics [19]. We used Voxceleb1 and Voxceleb2 in training the x-vector network and training the denoising techniques.

**Musan**: The Musan corpus has 109 hours of speech data with 60 hours of speech, 42 hours of music, and 929 noise files [20]. The Musan corpus is used for data augmentation in training x-vector systems.

**BBC Noise**: The BBC Noise corpus contains 16,000 sound effects [21]. The BBC noises are used as unseen noises and they are added to train and test x-vectors in denoising techniques.

**Fabiol**: Fabiol is a French corpus consisting of 6882 utterances. The length of files spans from very short utterances less than 2 seconds to very long utterances. The Faboil corpus is used for test and enrollment dataset.

### B. Train x-vector extractor

In our experiments, we used the standard Kaldi x-vector network introduced in [11]. The training data is augmented with different branches of Musan corpus (music, babble, noise, reverberation). Then, we extracted MFCC features for the augmented data. The MFCC features are normalized by Cepstral Mean Variance Normalization (CMVN) and silent frames are removed by the VAD. To explore the reliance of the x-vector system on the number of speakers and utterances in noisy environments, we trained two different networks. The first one is trained with 500,000 augmented utterances from Voxceleb1. In the second one, 1,000,000 randomly chosen utterances from Voxceleb1 and Voxceleb2 were used. In each protocol the trained network used to extract train and test x-vectors that used in denoising techniques.

### C. Train and test x-vectors used for denoising techniques

In denoising techniques, we need the pairs of noisy-clean x-vectors. We use two protocols to see the effectiveness of denoising techniques on x-vectors extracted from a network trained with poor data (Voxceleb1) and the network trained with more rich data (Voxceleb1 + Voxceleb2). The details about training and test dataset used in denoising techniques are described in the following.

**Protocol1:** In this protocol the x-vectors are extracted by the network trained with Voxceleb1. Firstly, the x-vectors for clean files in Voxceleb1 and Voxceleb2 are extracted. The BBC Noises and Musan were added to Voxceleb1 and Voxceleb2 with different SNRs from 0 to 15 to create the corresponding noisy x-vectors for each clean file. We used 1638 noise files from BBC corpus. The train data consists of 1.975 million pairs of noisy-clean x-vectors. It deserves to be mentioned that for some clean files there is more than one noisy version. For the test and enrollment dataset the Fabiol Corpus is used. The Fabiol corpus includes 6882 utterances that 3441 files were used for enrollment and the remaining part used as the test dataset. The test utterances corrupted with 547 different noises from BBC corpus with different SNRs between 0 and 15. Since the length of utterances in Fabiol is varied from very short (less than 2 seconds) to longer utterances (more than 12 seconds), we separated utterances by their duration in 6 groups to see the results of denoising methods on each group and specially to observe the effectiveness of the denoising techniques on very short utterances.

**Protocol 2:** In this protocol the x-vector network is trained with 1,000,000 from Voxceleb1 and Voxceleb2. The train dataset includes 1,200,000 pairs of noisy-clean vectors from Vxoceleb1 and Voxceleb2. The added noises are the same as protocol1. In this dataset for each noisy x-vector there is only one clean version. The test and enrollment files extracted by this network are the same as protocol 1.

## IV. Results

In this section, we describe the results of experiments for baseline system and denoising techniques. The results are briefed in Table 1 and Table 2. In the experiments, the equal error rate (EER) metric is used to evaluate the performance of the speaker recognition system. In all experiments, the PLDA classifier is used for scoring.

**Clean**: In this experiment the scoring is done on clean x-vectors in the test dataset. We can see that the results are strongly dependent on the duration of test files.

**Noisy**: To see the performance of x-vector system in noisy environments, the BBC noise files were added to the test data. In Table 1, we can see that there is a drastic degradation in our results. For example, in Protocol1 for utterances longer than 12 seconds the EER increased from 0.833 to 5.131. From Table 2 we can see that increasing the number of speakers and number of training data makes the system more robust but still there is a large drop in the performance of the system after adding the noise to the test data set. For example, in utterances longer than 12 seconds the EER increased from 0.5% to 2.69%.

**Denoising autoencoder:** Finding a good architecture and its parameters for a specific problem is the main challenge of using denoising autoencoders. In our experiments we used a denoising autoencoder with three layers. The input and output layers' activation function are linear. The hidden layer has 1024 neurons with *tanh* activation function. The network optimized by stochastic gradient descent algorithm. The learning rate was 0.02 that decays 0.0001 at each epoch. The network is trained in 100 epochs to reduce mean square error (MSE) loss function. We observed that in the case of using MSE loss, even the small improvement of MSE has a great impact on the results. In all experiments with conventional DAE and its modifications in the next experiments, we used Tensorflow [22] and Keras [23] frameworks. From Table 1 and Table 2 we can see that in all cases the denoising autoencoder improves the performance of the system in terms of EER. In Protocol1 we have 14% to 47% relative improvement of EER. This improvement in Protocol2 is from 19% for utterances less than 2 seconds to 58% for utterances between 8 and 10 seconds. The improvement for utterance longer than 10 seconds is 52%.

**Deep stacked denoising autoencoder:** In this experiment, we used deep dtacked denoising autoencoder. This architecture is described in section II. We used two DAE blocks. In the first one we put three layers. The input and output layer are linear and a dense layer with 1024 neurons were used in the hidden layer with *tanh* activation function. The output of the first DAE block concatenated with the difference between noisy x-vector and the output layer from the first DAE. This concatenated vector is used in the input of the next DAE block. In the second DAE, we used two *tanh* layers with 1024 neurons and the output layer is linear. The number of neurons in the output layer is 512 that is equal to the size of the noisy vector. The stochastic gradient descent optimization method is used to train the network. The learning rate is 0.02 and the decay of learning rate is 0.0001. From Table 1 and Table 2, we can see that in all experiments the stacked denoising autoencoder outperforms the denoising autoencoder. In Protocol1, we have 18% relative improvement for utterances shorter than 2 seconds and 51% improvement for utterances longer than 12 seconds. In Protocol2, we have 21% improvement for utterances shorter than 2 second and 66% improvement for utterances between 8 and 10 second. The results show that even with smaller number of training samples in denoising techniques the improvements in protocol 2 are higher. From this point we infer that training a good x-vector network results in more improvement by denoising techniques.

TABLE 1. The results for x-vectors extractor trained with Voxceleb1 (Protocol1) and denoising techniques

| Duration | s<2 | 2<s<4 | 4<s<6 | 6<s<8 | 8<s<10 | 10<s<12 | 12<s |
|---|---|---|---|---|---|---|---|
| Clean | 11.59 | 7.646 | 4.144 | 2.239 | 3.111 | 1.538 | 0.8339 |
| Noisy | 15.94 | 12.88 | 10.5 | 7.836 | 8.889 | 6.667 | 5.131 |
| DAE | 13.62 | 10.87 | 8.287 | 5.597 | 5.778 | 4.103 | 2.694 |
| **Stacked DAE** | **13.04** | **10.46** | **8.011** | **5.224** | **5.333** | **3.59** | **2.502** |

TABLE 2. The results for x-vector extractor trained with Voxceleb1+Voxceleb2 (Protocol2) and denoising techniques

| Duration | s<2 | 2<s<4 | 4<s<6 | 6<s<8 | 8<s<10 | 10<s<12 | 12<s |
|---|---|---|---|---|---|---|---|
| Clean | 10.43 | 4.628 | 1.934 | 1.119 | 0.888 | 1.026 | 0.577 |
| Noisy | 13.62 | 9.658 | 7.182 | 5.224 | 5.333 | 3.077 | 2.694 |
| DAE | 11.01 | 7.042 | 4.42 | 3.358 | 2.222 | 1.538 | 1.283 |
| **Stacked DAE** | **10.72** | **6.439** | **3.867** | **2.612** | **1.778** | **1.538** | **1.283** |

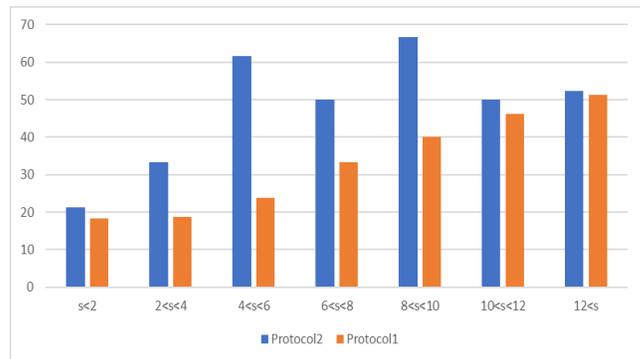

Fig 3. The relative improvement of EER(%) by deep stacked denoising autoencoder in Protocol1 and Protocol2

From the results in Table 1 and Table 2, we observed that, by increasing the number of speakers and using more data in training x-vector network, the system becomes more robust in deal with unseen noises. But the performance drops significantly in comparison to noise free environments. Applying noise compensation techniques brings notable improvement in both protocols. The relative improvement of EER for both protocols is presented in Fig 3.

## V. Conclusion

In this paper we proposed a modification in standard x-vector system to make the system more robust in noisy environments. The modification is adding a denoising subsystem before scoring x-vectors. As other studies showed, we conclude that the system's performance continue to improve by increasing the number of speakers and data argumentation. We showed also that even with this fact, applying compensation techniques are essential to approach free-noise test conditions.

### ACKNOWLEDGEMENT

This work has been supported by ROBOVOX and VoicePersonae ANR projects.